\documentclass[%
 reprint,
 amsmath,amssymb,
 aps,
 prd,
floatfix,
showkeys
]{revtex4-1}

\usepackage[USenglish]{babel}
\usepackage{slashed}
\usepackage{graphicx}
\usepackage{dcolumn}
\usepackage{bm}
\usepackage{amssymb}
\usepackage{amsmath}
\usepackage{graphicx}
\usepackage{epsfig}
\usepackage{rotating}

\usepackage[utf8]{inputenc} 
\usepackage{amsmath} 
\usepackage{array}
\usepackage{float}
\usepackage{color}
\usepackage[usenames,dvipsnames]{xcolor}
\usepackage{epstopdf}
\usepackage{natbib}
\usepackage{graphicx}
\usepackage{lineno}
\usepackage{slashed}
\usepackage{color}
\usepackage{hyperref}
\usepackage{float}
\usepackage{mathtools,slashed}
\usepackage{multirow}

\newcommand{\be}{\begin{equation}}
\newcommand{\ee}{\end{equation}}
\newcommand{\bea}{\begin{eqnarray}}
\newcommand{\eea}{\end{eqnarray}}
\newcommand{\beas}{\begin{eqnarray*}}
\newcommand{\eeas}{\end{eqnarray*}}

\newcommand{\bse}{\begin{subequations}}
\newcommand{\ese}{\end{subequations}}

\begin{document}

\title{\bf Differential cross section for elastic collision, first exited state and ionization of  water molecule}
\author{C. H.~Zepeda~Fern\'andez$^{1,2}$}
\email{Corresponding author, email:hzepeda@fcfm.buap.mx}
\author{U. Reyes~Ortega$^{3}$}
\author{F. Maga\~na Iba\~nez$^{2}$}
\author{E.~Moreno~Barbosa$^2$}

\address{
  $^1$Cátedra CONACyT, 03940, CdMx Mexico\\
  $^2$Facultad de Ciencias F\'isico Matem\'aticas, Benem\'erita Universidad Aut\'onoma de Puebla, Av. San Claudio y 18 Sur, Ciudad Universitaria 72570, Puebla, Mexico\\
  $^3$Facultad de F\'isica, Universidad Veracruzana, Lomas del Estadio, 91090 Xalapa-Enríquez, Veracruz.
}

\begin{abstract}
\begin{center}
\bf \large Abstract
\end{center}
The radiolysis of water is an important study in microdosimetry, one of the resulting and important molecules in this study is the $OH^-$. Theoretical quantum mechanics calculations  are used to obtain the differential cross section for the elastic collision, first excited state and ionization of water molecule. For the ionization case, obtaining  $OH^-$ is studied. As incident particles are  considered electrons and protons, independently for each phenomena previously mentioned. The differential cross section was obtained by the amplitude function. The water molecule model used to calculations consisted by the two protons of the hydrogen atom that make up the water molecule and the structure-less and doubly charged oxygen. For the proton incident particle case, the identical particle analysis was used. The results show the angular dependence and therefore the interference phenomenon between the incident wave and the deflected wave.
\end{abstract}

\keywords{Differential cross section, water molecule, elastic collision, inelastic collision.}

\maketitle

\section{Introduction}\label{1}
One of the most important phenomena for the interaction of radiation and water molecule is in the biological field. For example, the human body is mainly composed of water, the 80\% of human tissues consist of water molecule~\cite{water}. Then, the water molecule is the most important molecule used in Monte Carlo to model cells and organs. One of the most important studies are to know the differential cross section or total cross section of the water radiolysis~\cite{Rad, MC1,MC2,MC3,MC4}, being one of the most important diuretic effects of the radioactivity. This phenomenon is complex, it can be summarized as the breaking of the water molecule and  therefore formation of free radicals. The chemical species produced in the radiolysis are radioactive and modifies biological molecules, which can turn into cancer~\cite{rad}. In principle, the incident particle  interacts with all the components of the water molecule, then it is necessary to have a model. Modeling the water molecule is nearly impossible, because the Schr\"odinger equation would have to be solved for a system of many bodies and interacting with each other (28 particles: 10 protons, 8 neutrons and 10 electrons). Various studies have been carried out since long ago to find the wave function and quantum numbers using either suitable approximations~\cite{wm1,wm2,wm3,wm4} and numerical methods~\cite{wm5,wm6}.\\
In some many studies have been calculated the cross section of the interaction of protons with water molecule, considering a model of electronic structure~\cite{elestruct} and considering electrons as incident particles~\cite{diffelectron}.\\
One of the molecules obtained in the radiolysis of water is the $OH^-$, which it is obtained by the ejection of a nucleus of a hydrogen atom. In this work, it is used a simple model for the water molecule, based on three particles system, two protons from the hydrogen atom and the oxygen doubly charged and unstructured, to calculate the differential cross section ($d\sigma/d\Omega$) for an elastic collision, the first exited state and ionization of water molecule. For each phenomena, were considered as incident particles protons and electrons. In total six results were obtained. The work is organized as follows: In Section~\ref{model} is shown the model used for the water molecule, as well as, its wave function. In  Section~\ref{crossfunction} it is found the amplitude function for the three phenomena and the respective  differential cross section. In Section~\ref{results} it is shown the differential cross section distribution for the two cases of incident particles  and finally in Section\ref{conclusions} the discuss and conclude are shown. 

\section{Water molecule model}\label{model}
The water molecule is a system of many interacting particles, then, finding an analytical solution of the Scr\"odinger's equation is nearly impossible. The studies has been achieved only after resorting to some simplifying assumptions. These studies have been made since long ago~\cite{wm7,wm8,wm9,wm10}. Also, the numerical methods have been used as an alternative~\cite{wm11,wm12}.\\
One of the simple models of the water molecule is to consider the two protons of the hydrogen atoms and the oxygen double negatively charged (due to the two electrons covalently bonded) and unstructured. Owing to the electrostatic interaction, it is known that the distance between the protons is constant ($d_p\sim$151.05~pm), as well as, the distance between the protons and the oxygen nucleus ($d_o\sim$95.60~pm)~\cite{we1}. 
This model has been studied to obtain its wave function~\cite{model}, which is giving by:
\begin{equation}
\Psi_{l_{1},l_{2},m_{1},m_{2}}(\theta_1,\phi_1;\theta_2,\phi_2)=Y_{l_1m_1}(\theta_1,\phi_1)Y_{l_2m_2}(\theta_2,\phi_2)
\end{equation}
where the subscript 1 and 2 represents each proton. $Y_{l_im_i}(\theta_i,\phi_i)$, are the spherical harmonic, $l_i$ and $m_i$ are the orbital angular and magnetic quantum number, respectively, for for $i=1,2$.

\section{Amplitude function and differential cross section}\label{crossfunction}
In a scattering experiment, the detectors are located at much greater distance than the target range interaction, then, it is well known that in this asymptotic limit, the amplitude function is giving by:

\begin{equation}\label{amplitude}
\displaystyle f(\theta,\phi)=-\frac{\mu}{2\pi\hbar^2}<\Psi_i|\hat V|\Psi_f>
\end{equation}

Where $\mu$ is the reduced mass of the system,  $|\Psi_i>$ is the initial state of the system before the collision, $|\Psi_f>$ is the final state of the system after the collision and $\hat V$ is the interaction potential. The differential cross section is related to the amplitude function by:
\begin{equation}\label{diff_cross}
    \displaystyle \frac{d\sigma}{d\Omega}=\frac{k_f}{k_i}|f(\theta,\phi)|^2
\end{equation}
where $k_i^2=2m_eE_i/\hbar^2$ and $k_f^2=2m_eE_f/\hbar^2$ are the wave number for the incident and scattering wave function, depending on the initial energy $E_i$ and final energy $E_f$,  respectively.\\
The initial state is depending on the incident particle, which is represented by a plane wave $\psi_{i}(\vec r)=e^{i\vec k_i \cdot \vec r}$.  Initially, the water molecule is in the ground state, then, the spatial wave function is giving by  $\Psi_{0,0,0,0}(\theta_1,\phi_1;\theta_2,\phi_2)=Y_{00}(\theta_1,\phi_1)Y_{00}(\theta_2,\phi_2)=\frac{1}{4\pi}$. Therefore, the initial state of the system is giving by:

\begin{equation}
\displaystyle \Psi_i(\vec r;\theta_1,\phi_1;\theta_2,\phi_2)=\frac{1}{4\pi}e^{i\vec k_i \cdot \vec r}
\end{equation}
The interaction between the incident particle and the oxygen is a Coulomb interaction $\pm \frac{2e^2}{r}$, where the $\pm$ represents an attractive or repulsive interaction, depending on the charge of the incident particle. The interaction with the protons also is a Coulomb potential $\mp\frac{e^2}{|\vec r-\vec d_{o}^1|}\mp\frac{e^2}{|\vec r-\vec d_{o}^2|}$, where the $\mp$ is depending on the charge of the incident particle. $\vec d_o^1$ and $\vec d_o^2$ are the distance between the oxygen and proton 1 and 2, respectively. The total potential is giving by,

\begin{equation}
\displaystyle \hat V= \pm\frac{2e^2}{r}\mp\frac{e^2}{|\vec r-\vec d_o^1|}\mp\frac{e^2}{|\vec r\mp\vec d_{o}^2|}
\end{equation}

Remember that $|\vec d_o^1|=|\vec d_o^2|=d_o$. Therefore, the Eq.~\ref{amplitude} becomes

\begin{equation}
\begin{split}
f(\theta,\phi)=&-\frac{\mu}{2\pi\hbar^2} \int \int \int \frac{1}{4\pi}e^{-i\vec k_i \cdot \vec r} \Psi_f\\ 
&\times \Big(\frac{2e^2}{r}-\frac{e^2}{|\vec r-\vec d_o^1|}-\frac{e^2}{|\vec r-\vec d_{o}^2|}\Big)  d\Omega_1 d\Omega_2 d^3r.
\end{split}
\end{equation}

$\Psi_f$ is the final state of the system, which depends of  the scattered particle and the final state of the water molecule. The scattered particle is detected far from the the interaction range of the potential, then, the wave function is described also as a plane wave $\psi_{s}(\vec r)=e^{i\vec k_f \cdot \vec r}$.\\
From Eq.~\ref{diff_cross}, the differential cross section is giving by, 

\begin{equation}\label{cross}
    \begin{split}
        \displaystyle \frac{d\sigma}{d\Omega}=&\frac{k_f}{k_i}|f(\theta,\phi)|^2\\
        =&\frac{k_f}{k_i}\Big|-\frac{\mu}{2\pi\hbar^2} \int \int \int \frac{1}{4\pi}e^{i\vec k_i \cdot \vec r'} \Psi_f\\ 
&\times \Big(\pm\frac{2e^2}{r}\mp\frac{e^2}{|\vec r\mp\vec d_o^1|}\mp\frac{e^2}{|\vec r-\vec d_{o}^2|}\Big)  d\Omega_1 d\Omega_2 d^3r\Big|^2.\\
    \end{split}
\end{equation}

In the subsections below, Eq~\ref{cross} is applied when the final state of the water molecule is elastic collision, first excited state and the ionization ($OH^-$ molecule). Before continuing with the results, it is necessary to remember the analysis required for the case of scattering of identical particles, i.e., when protons are chosen as incident particles.

\subsection{Scattering of identical particles}
For the scattering of two identical particles with spin $1/2$, the system is known to be either symmetric or anti-symmetric. From the Quantum Mechanics scatter theory, when the two particles are in an anti-symmetric state, that is, the two particles are in a singlet state, the Eq.~\ref{cross} becomes

\begin{equation}\label{singlet}
\frac{d\sigma_S}{d\Omega}=\frac{k_f}{k_i}|f(\theta)+f(\pi -\theta)|^2.
\end{equation}
For the other hand, when the two particles are in a spin triplet state, the wave function is anti-symmetric, then the Eq.~\ref{cross} becomes 
\begin{equation}\label{triplet}
\frac{d\sigma_T}{d\Omega}=\frac{k_f}{k_i}|f(\theta)-f(\theta-\pi)|^2.
\end{equation}

Finally, when the particles are unpolarized, the states are equally likely, so the triplet state is three times as likely as the singlet, then,

\begin{equation}\label{all}
\frac{d\sigma}{d\Omega}=\frac{3}{4}\frac{d\sigma_T}{d\Omega}+\frac{1}{4}\frac{d\sigma_S}{d\Omega}.
\end{equation}
Eq~\ref{singlet}, Eq~\ref{triplet} and Eq~\ref{all} are used for proton scattering.

\section{Analysis and results}\label{results}

\subsection{Elastic scattering of water molecule}
\subsubsection{Electrons as incident particles}
For this case, the incident electron, with energy $E_i$ does not deposit energy, then, $k_i=k_f$ (or $E_i=E_f$). The final wave function for the water molecule is the same as the initial, therefore, the final state of the system is $\Psi_f(\vec r)=\frac{1}{4\pi}e^{i\vec k_f \cdot r}$ and the amplitude function becomes,

\begin{equation}\label{f_elastic}
    \begin{split}
f_E(\theta,\phi)=&-\frac{\mu}{2\pi\hbar^2}\frac{1}{(4\pi)^2} \int \int \int e^{i\vec q_E \cdot \vec r}\\ 
&\times \Big(\frac{2e^2}{r}-\frac{e^2}{|\vec r-\vec d_o^1|}-\frac{e^2}{|\vec r-\vec d_{o}^2|}\Big)  d\Omega_1 d\Omega_2 d^3r.
\end{split}
\end{equation}

\begin{figure}[htbp]
\begin{center}
\includegraphics[width=0.5\textwidth]{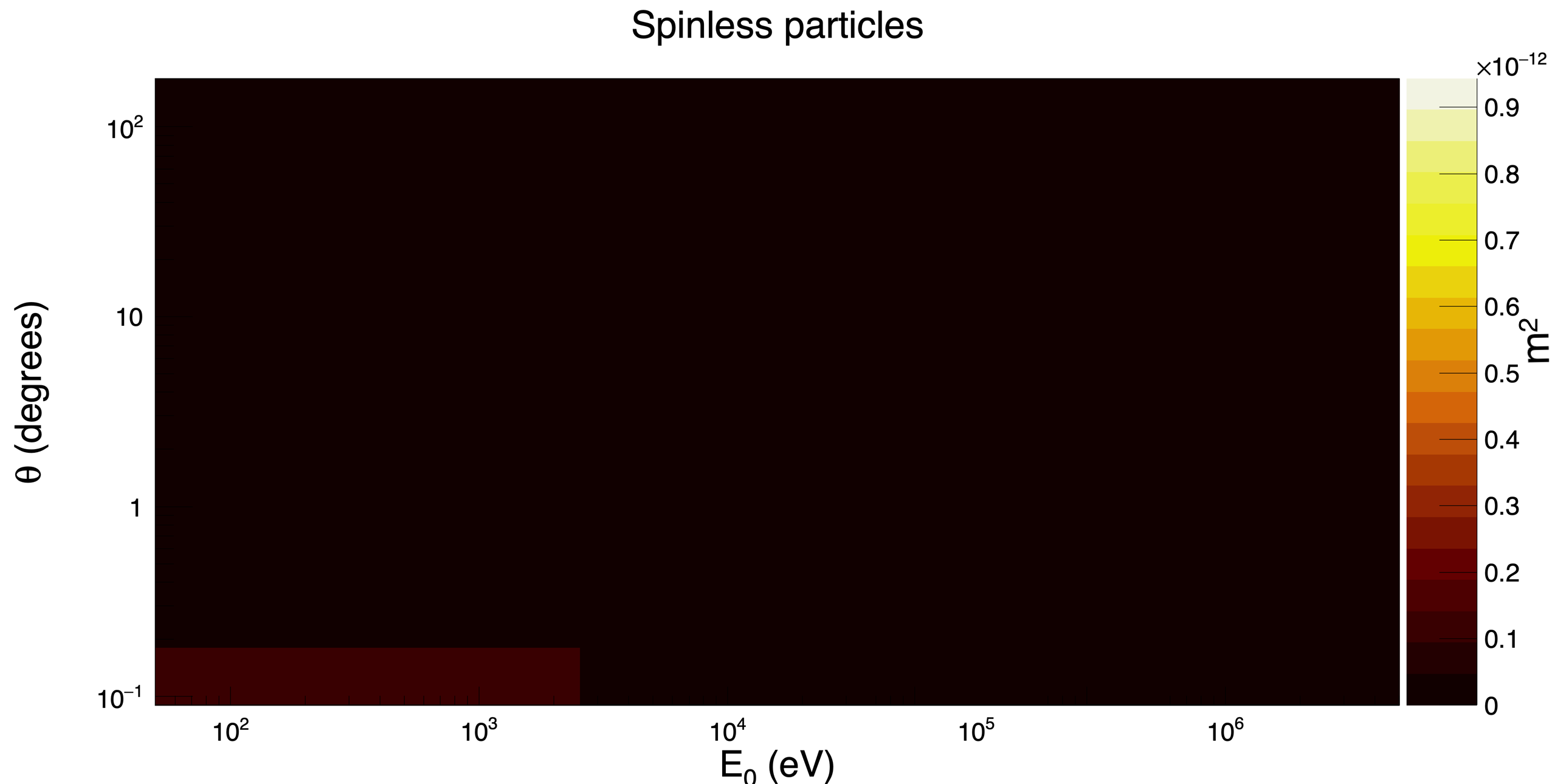}
\end{center}
\caption{Elastic collision correlation for $d\sigma_E/d\Omega$  between the scatter angle and kinetic energy for electrons.}
\label{electron_map_elastic}
\end{figure}

\begin{figure}[htbp]
\begin{center}
\includegraphics[width=0.5\textwidth]{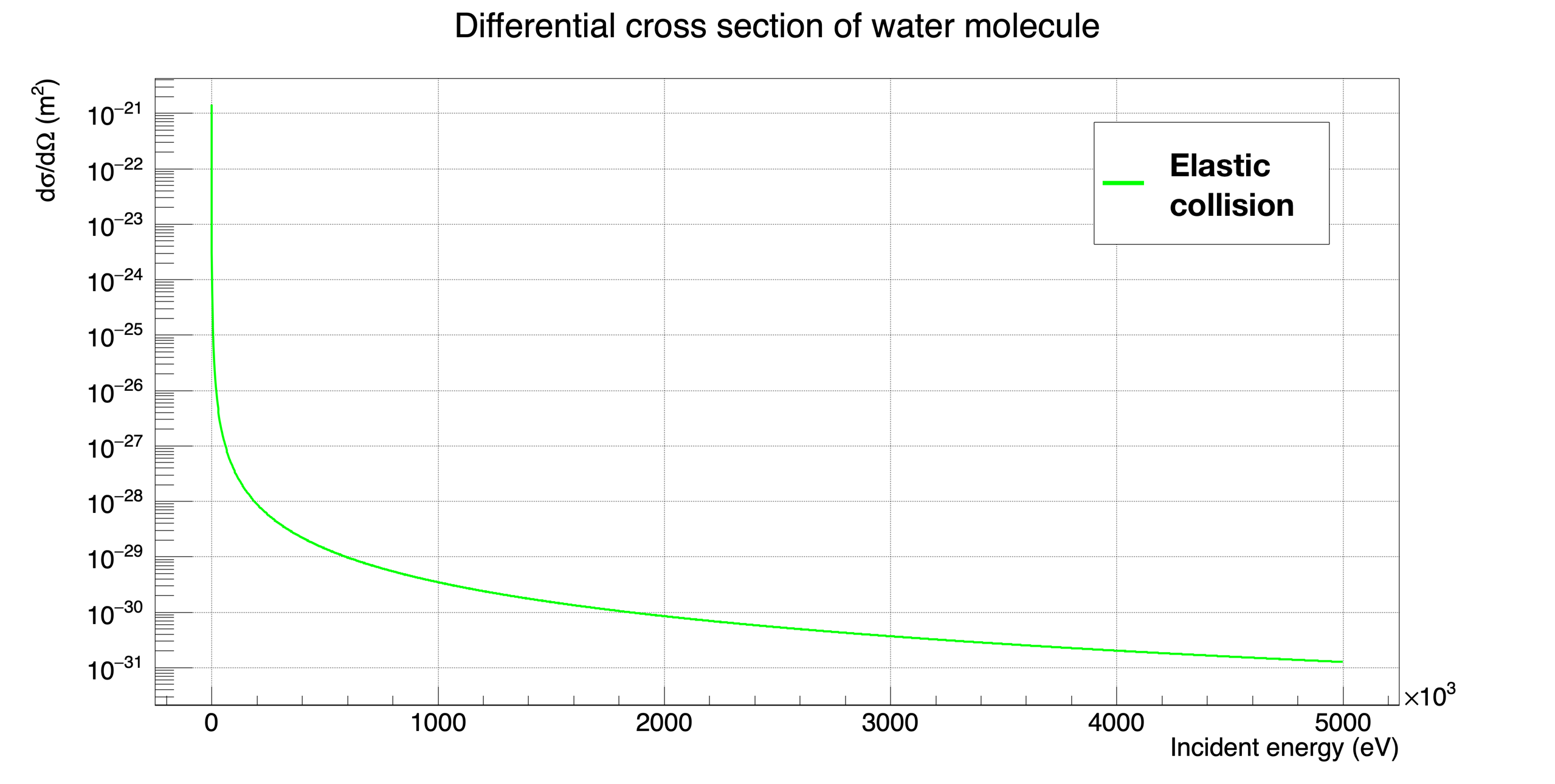}
\end{center}
\caption{Differential cross section for a scatter angle at 45$^o$, due to an elastic collision of electrons, as function of the kinetic energy.}
\label{electron_dif_elastic_fixangle}
\end{figure}

\begin{figure}[htbp]
\begin{center}
\includegraphics[width=0.5\textwidth]{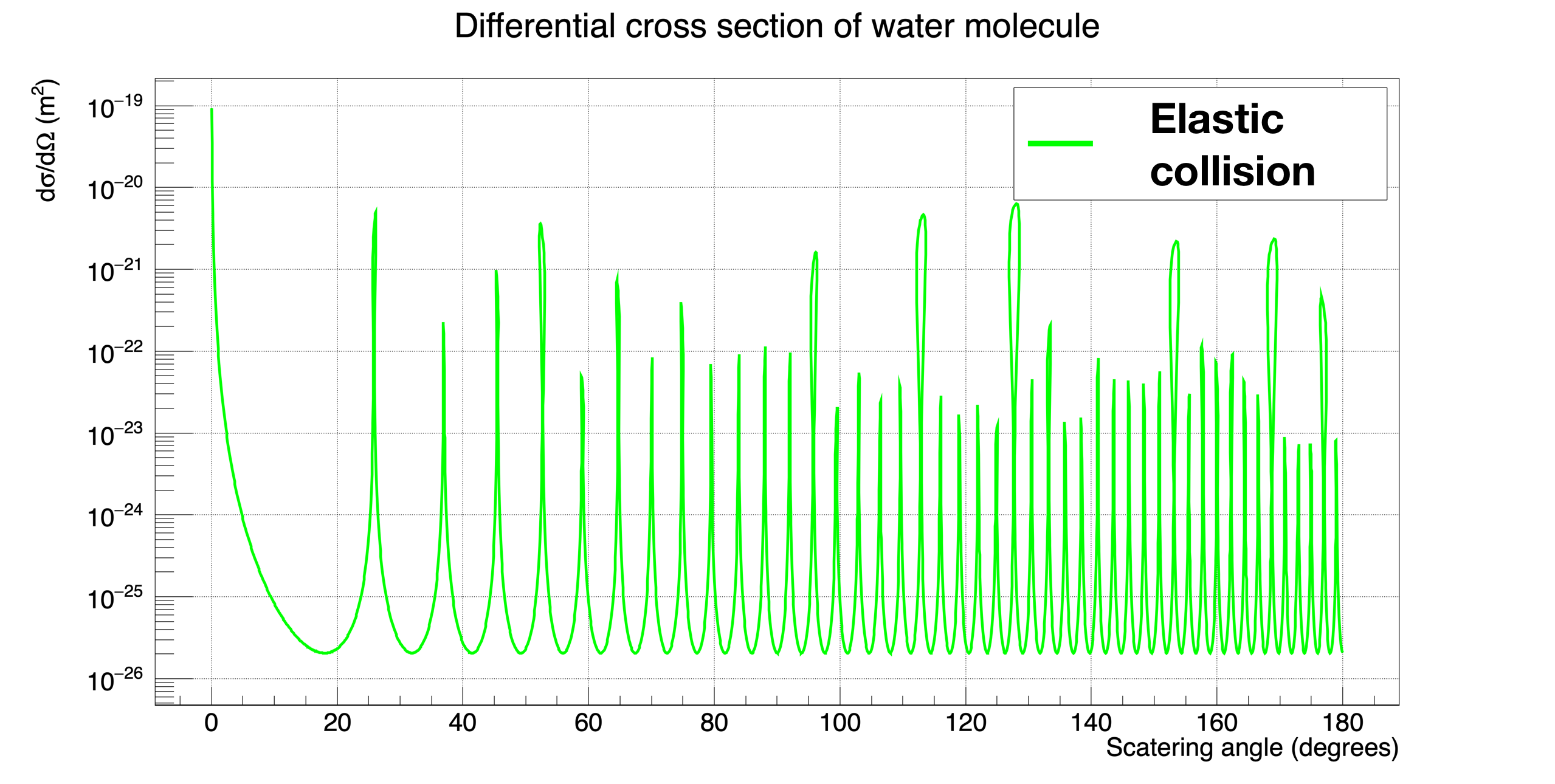}
\end{center}
\caption{Differential cross section for an  incident electron of 500~keV, due to an elastic collision, as function of the scatter angle.}
\label{electron_dif_elastic_fixE}
\end{figure}

Where $\vec q_E= \vec k_f-\vec k_i$, $d\Omega_1$ and $d\Omega_2$ are the solid angle differential for the protons of the water molecule and $d^3r$ is the volume differential for the incident electron. The integrals can be solved straight forward using $\int e^{i\vec q \cdot \vec r}/rd^3r=4\pi/q^2$, then the differential cross section is giving by

\begin{equation}\label{diff_elastic}
\displaystyle \frac{d\sigma_E}{d\Omega}= |f_E(\theta,\phi)|^2=\Big(\frac{\mu}{2\pi\hbar^2}\Big)^2\frac{4}{q_E^4} \Big[1+\frac{sin(q_Ed_0)}{q_Ed_0}\Big]^2
\end{equation}

where $q_E=2\sqrt{\frac{2\mu E_i}{\hbar^2}}sin\big(\frac{\theta}{2}\big)$. In Figure~\ref{electron_map_elastic} is shown the correlation between the scatter angle with the incident electron kinetic energy. As particular cases, in Figures~\ref{electron_dif_elastic_fixangle} and \ref{electron_dif_elastic_fixE} are shown the values of Eq.~\ref{diff_elastic} for a fixed angle of 45$^o$ and the initial kinetic energy of 500~keV, respectively.

\subsubsection{Protons as incident particles}
\begin{figure}[htbp]
\begin{center}
\includegraphics[width=0.5\textwidth]{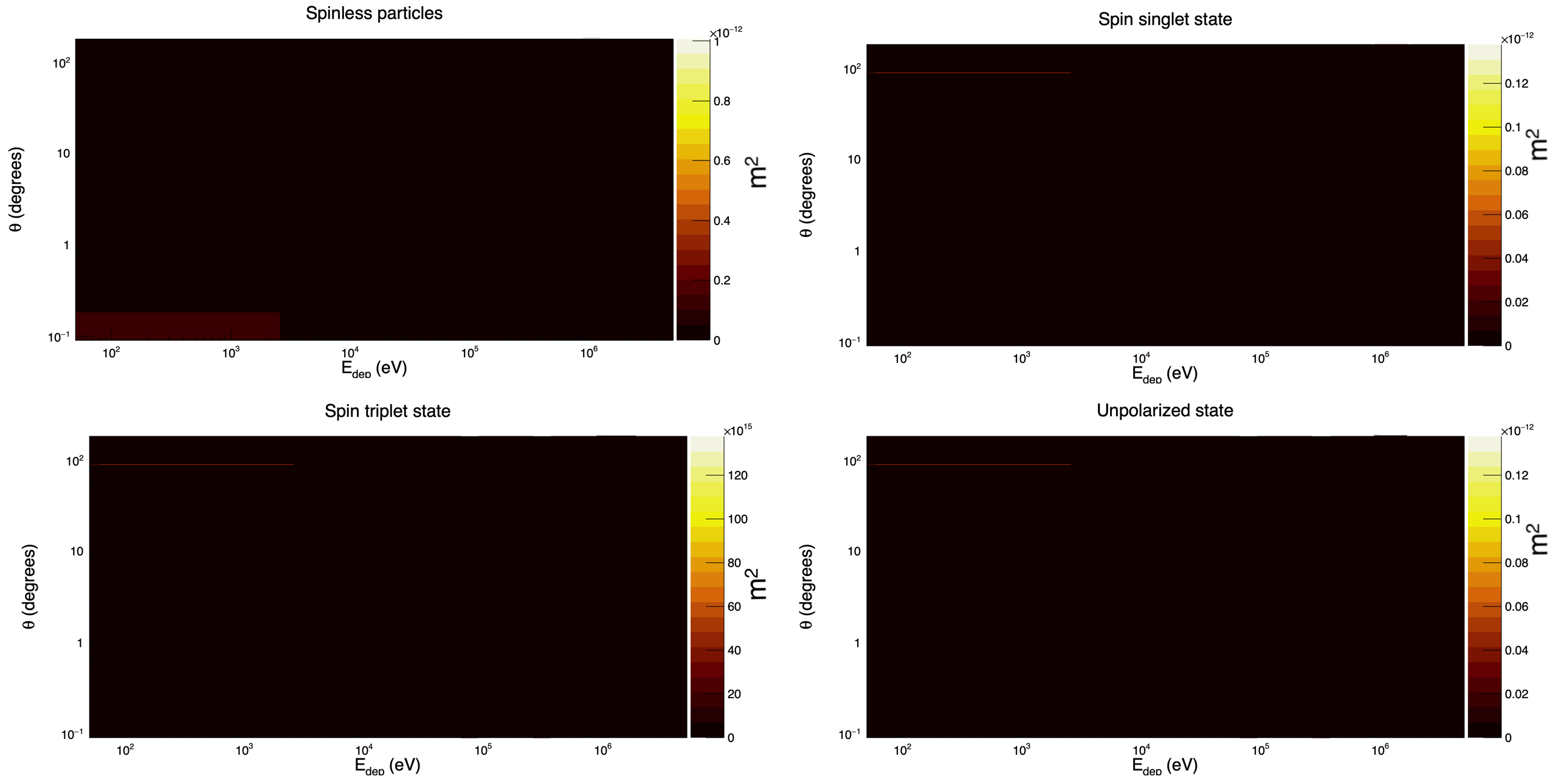}
\end{center}
\caption{Correlation for $d\sigma/d\Omega$  between the scatter angle and kinetic energy for elastically scattered protons. In the top left corner is shown the results for the case without spin, the top right corner is for spin singlet state, the bottom left corner shows the triplet case and the right bottom corner shows the results for the unpolarized state.}
\label{proton_map_elastic}
\end{figure}

\begin{figure}[htbp]
\begin{center}
\includegraphics[width=0.5\textwidth]{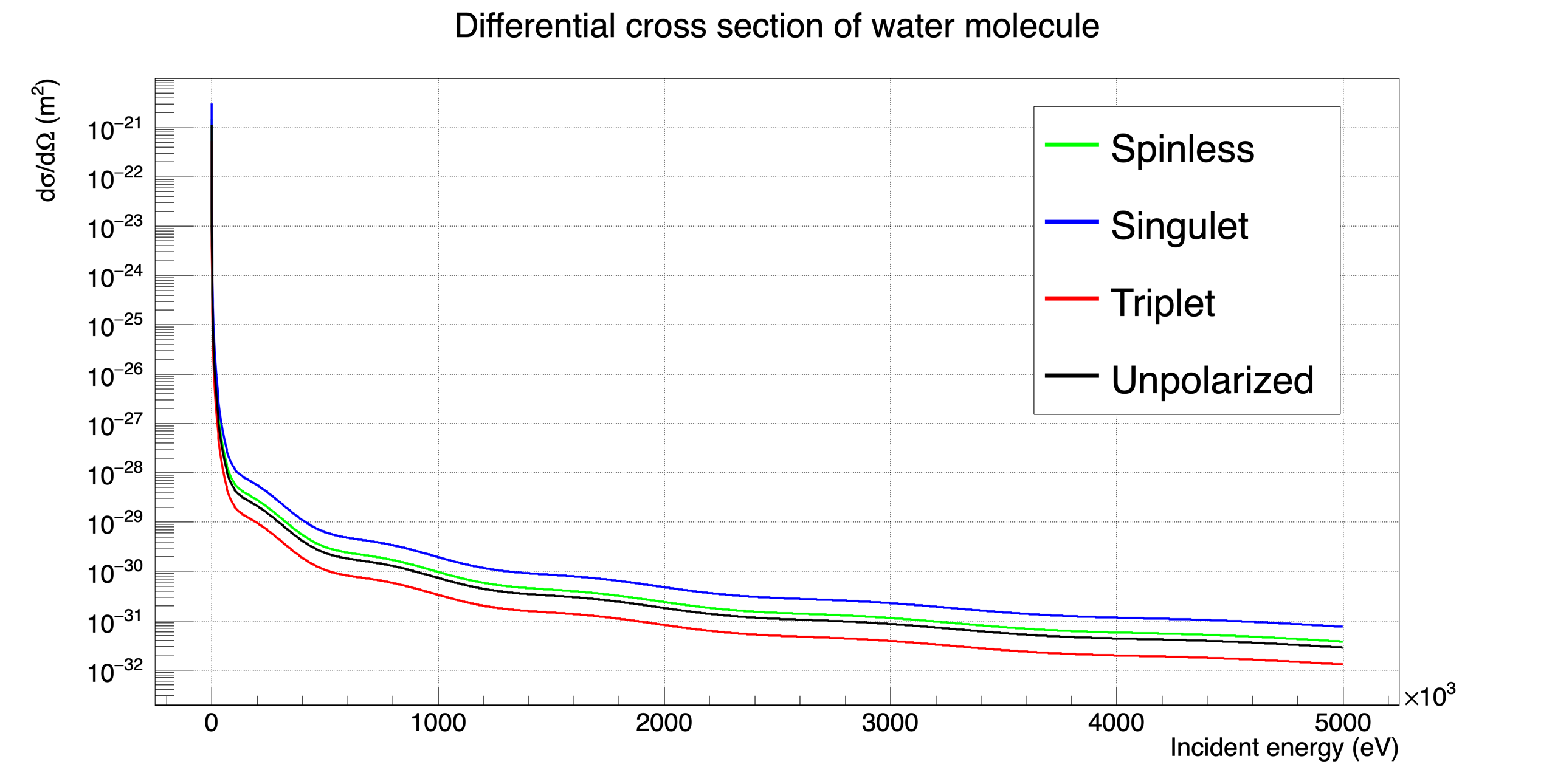}
\end{center}
\caption{Differential cross section for a scatter angle of 45$^o$, due to an elastic collision of protons, as function of the kinetic energy.}
\label{proton_dif_elastic_fixangle}
\end{figure}

\begin{figure}[htbp]
\begin{center}
\includegraphics[width=0.5\textwidth]{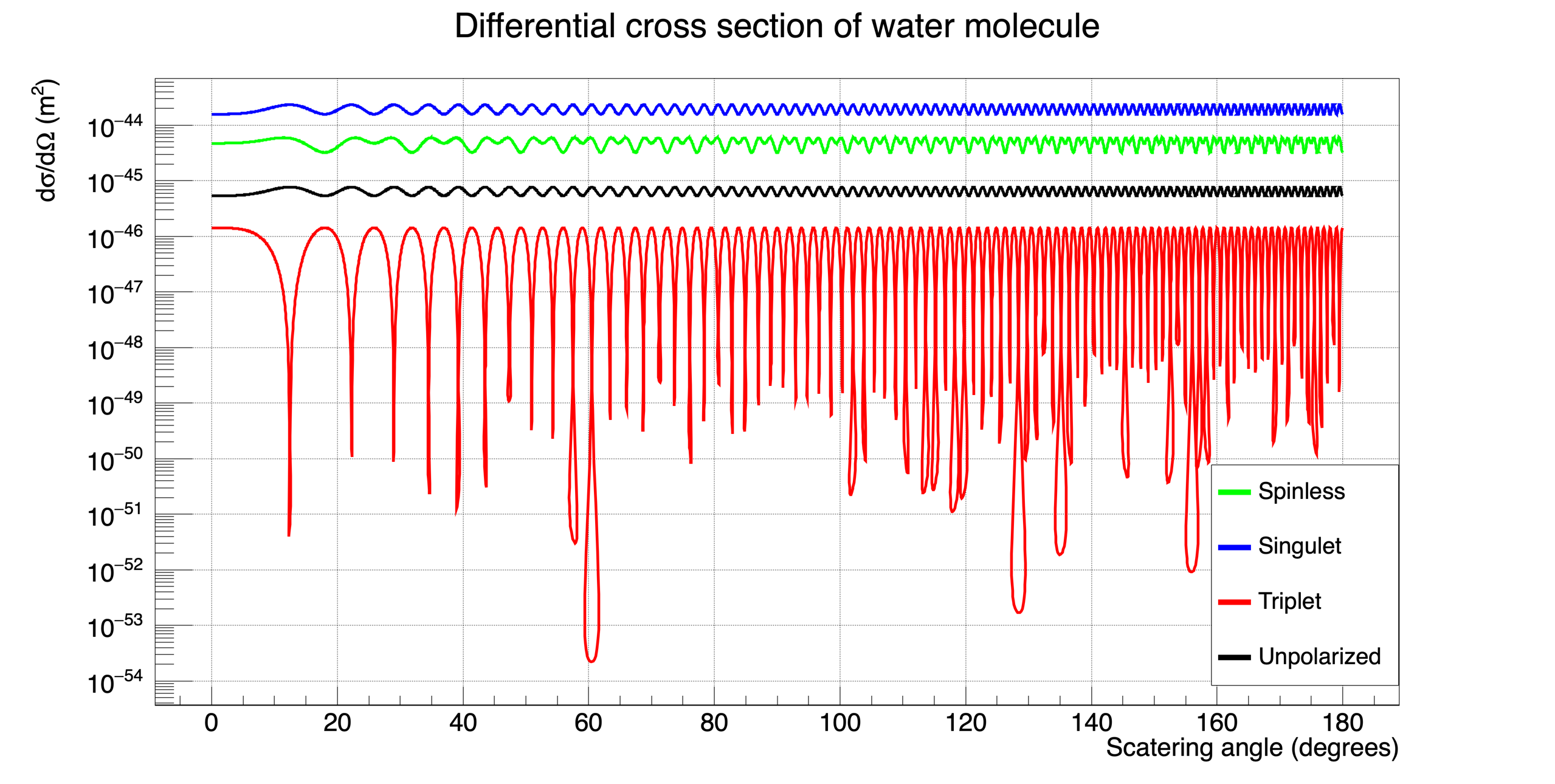}
\end{center}
\caption{Differential cross section for incident protons of 500~keV, due to an elastic collision with a water molecule, as function of the scatter angle.}
\label{proton_dif_elastic_fixE}
\end{figure}

For the case of scattering protons, there is a system of identical particles, then, the protons can be in a singlet or a triplet state. The amplitude function is giving as Eq~\ref{f_elastic}, by reversing the signs as mentioned in Eq.~\ref{cross}. However, the differential cross section is the same as  Eq.~\ref{diff_elastic}, due to Eq.~\ref{diff_cross}. In Figure~\ref{proton_map_elastic} is shown the correlation between the scatter angle and the incident energy to consider spinless, singlet state, triplet state and not polarized state. Finally, in  Figures~\ref{proton_dif_elastic_fixangle} and \ref{proton_dif_elastic_fixE}  are shown the values of Eq.~\ref{diff_elastic} for a fixed angle of 45$^o$ and an initial kinetic energy of 500~keV, respectively. 

\subsection{First excited state of water molecule}
According to~\cite{model} the first exited state of the water molecule is giving when the spatial function of one of the protons is $Y_{1m}(\theta,\phi)$, where $m=-1,0,1$. It is considered the proton 2 located in the first excited state,  then,  $l_2=1$, and its state functions are giving by:

\begin{equation}
\begin{split}
\displaystyle & Y_{10}(\theta_2,\phi_2)=\sqrt{\frac{3}{4\pi}}cos(\theta_2)\\               & Y_{1\pm1}(\theta_2,\phi_2)=\mp\sqrt{\frac{3}{8\pi}}e^{\pm i\phi_2}sin(\theta_2)
\end{split}
\end{equation}

Due to the symmetry of the potential, the integral in $\phi_2$ coordinate vanish for $Y_{1\pm1}(\theta_2,\phi_2)$, therefore, the only exited state is giving by $Y_{10}(\theta_2,\phi_2)$. The final state of the system is $\Psi_f(\theta_2,\vec r)=\frac{1}{\sqrt{4\pi}}\sqrt{\frac{3}{4\pi}}cos(\theta_2)e^{-i\vec k_f \cdot \vec r}$. Then, from Eq.~\ref{cross} it is obtained,

\begin{equation}
    \begin{split}
        \displaystyle \frac{d\sigma_{fes}}{d\Omega}=&\frac{k_f}{k_i}|f(\theta,\phi)|^2\\
        =&\frac{k_f}{k_i}\Big|-\frac{\mu}{2\pi\hbar^2}\sqrt{\frac{3}{4\pi}}\frac{1}{\sqrt{4\pi}} \int \int \int e^{i\vec q_{fes} \cdot \vec r}cos(\theta_2)\\ 
&\times \Big(\frac{2e^2}{r}-\frac{e^2}{|\vec r-\vec d_o^1|}-\frac{e^2}{|\vec r-\vec d_{o}^2|}\Big)  d\Omega_1 d\Omega_2 d^3r\Big|^2.\\
    \end{split}
\end{equation}

Where $\displaystyle q_{fes}=|\vec k_f-\vec k_i|$ and for this case $k_f\neq k_i$. The integrals for the $\frac{2e^2}{r}$ and $-\frac{e^2}{|\vec r-\vec d_o^1|}$
 potential contributions are zero, were there is a term of the form $\displaystyle \int_0^\pi cos(\theta_2)sin(\theta_2)=0$. The only contribution is giving by the third term. The final result is
 
 \begin{equation}\label{diff_exc}
     \displaystyle \frac{d\sigma}{d\Omega}=12\frac{\mu^2e^4}{\hbar^4q_{fes}^8d_0^4}\frac{k_f}{k_i}\Big[sin(q_{fes}d_o)-q_{fes}d_ocos(q_{fes}d_o)\Big]^2
 \end{equation}
 
 \subsubsection{Electrons as incident particles}
\begin{figure}[htbp]
\begin{center}
\includegraphics[width=0.5\textwidth]{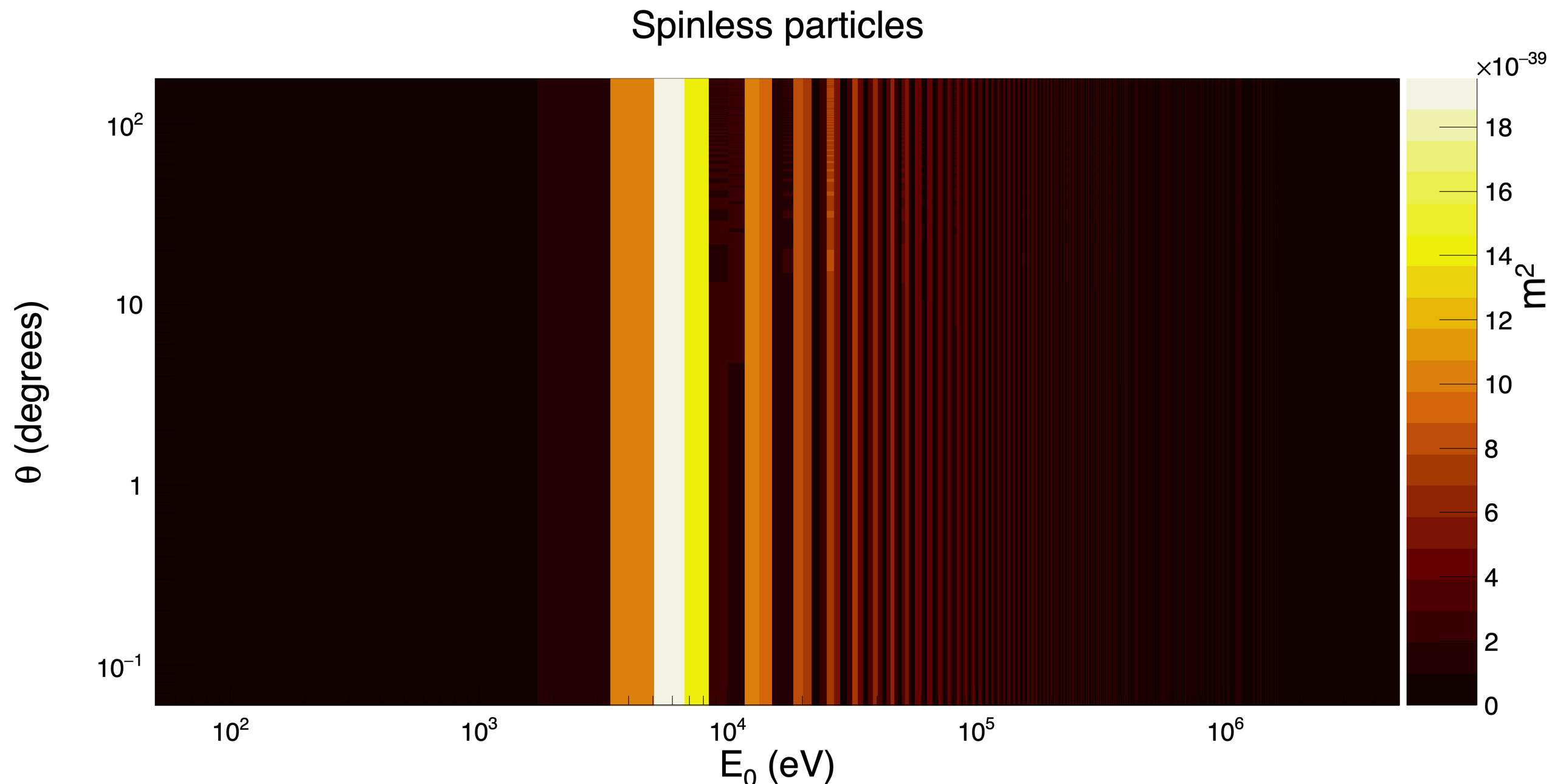}
\end{center}
\caption{Correlation for $d\sigma/d\Omega$  between the scatter angle and kinetic energy for electrons as incident particles, when the excitation of the water molecule occurs.}
\label{electron_map_excit}
\end{figure}

\begin{figure}[htbp]
\begin{center}
\includegraphics[width=0.5\textwidth]{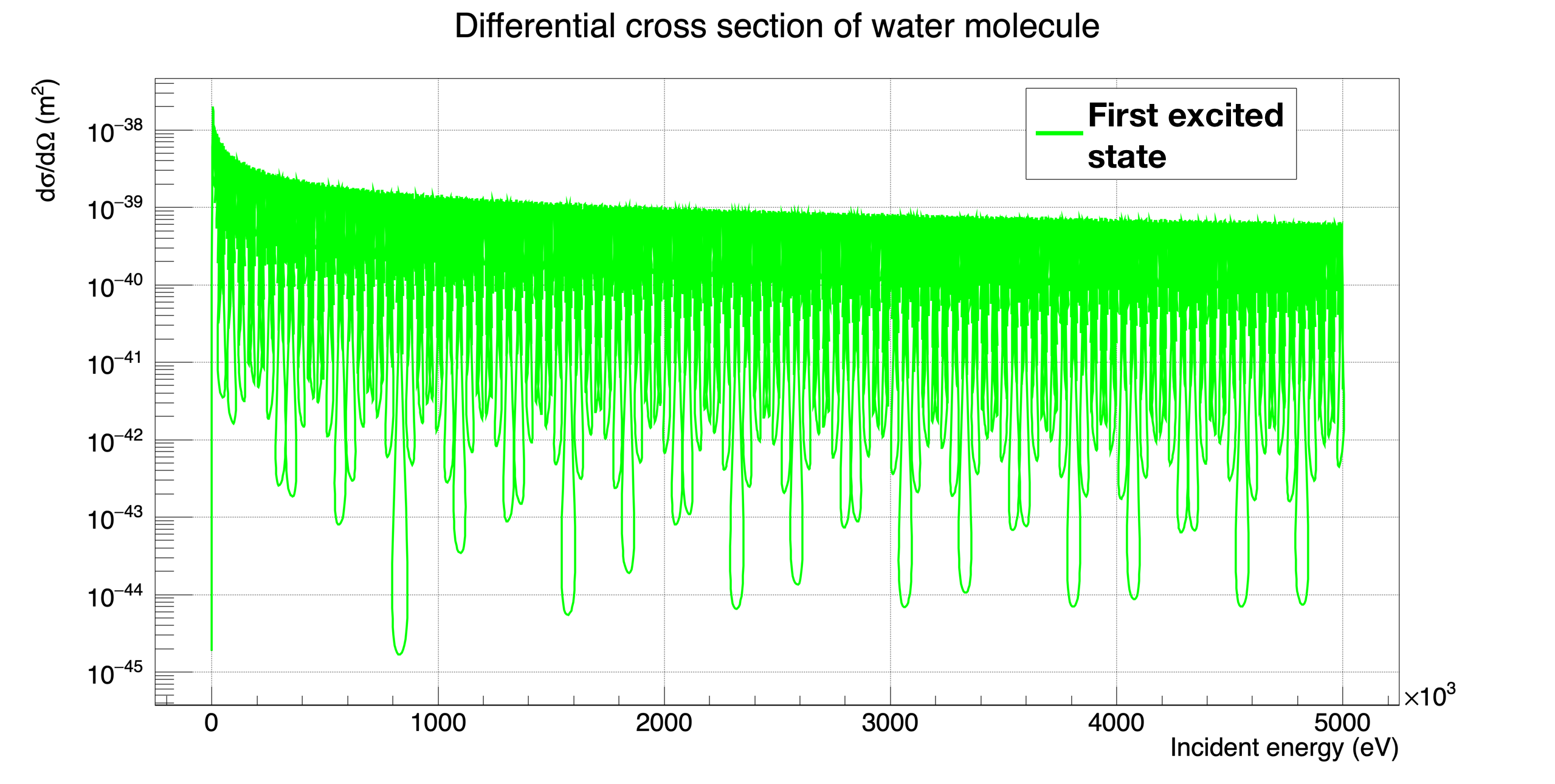}
\end{center}
\caption{Differential cross section for a scatter angle of 45$^o$ as function of the kinetic energy of electrons, when the water molecule excitation occurs.}
\label{electron_dif_excit_fixangle}
\end{figure}

\begin{figure}[htbp]
\begin{center}
\includegraphics[width=0.5\textwidth]{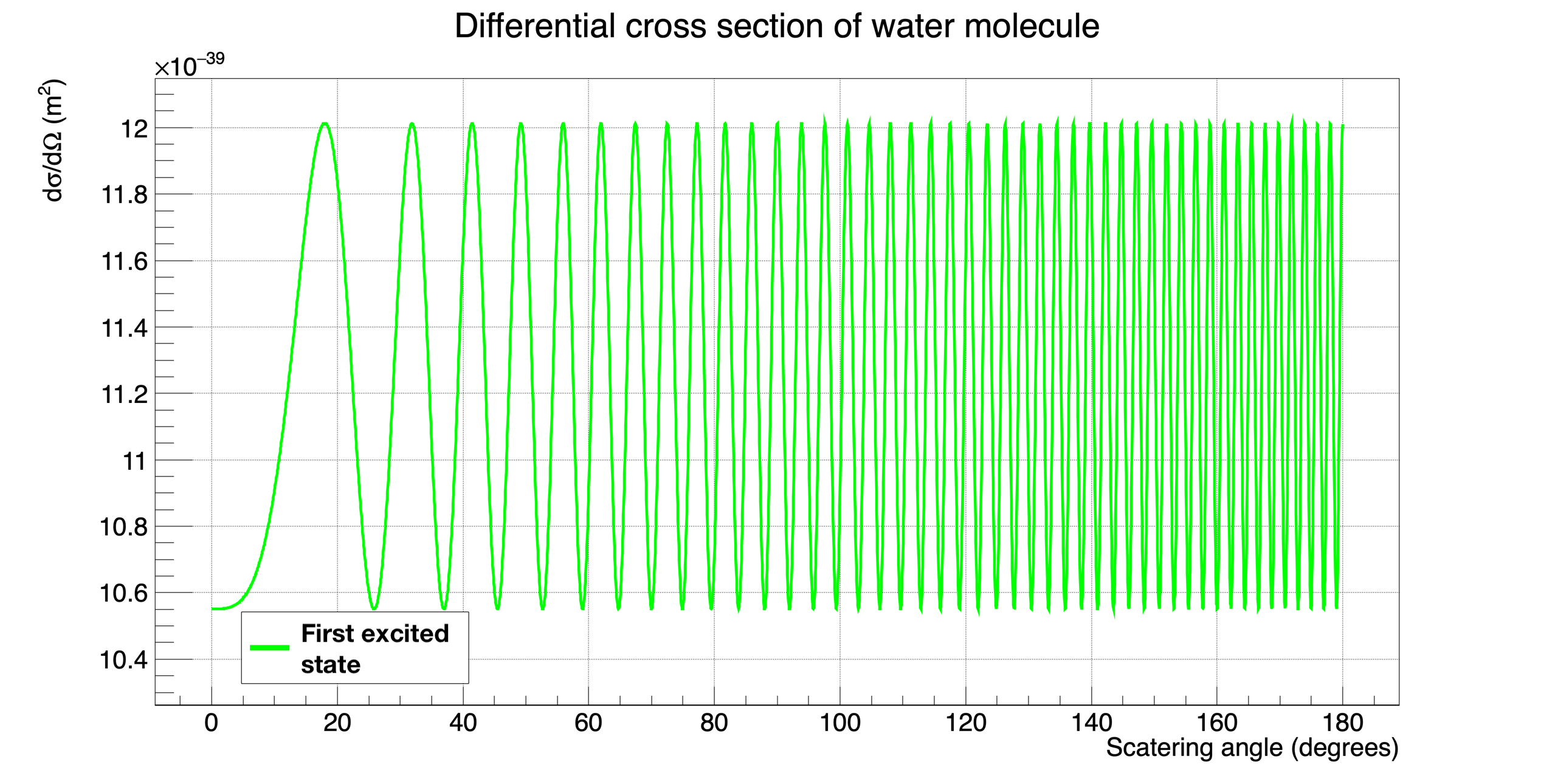}
\end{center}
\caption{Differential cross section for the   excitation of water molecule for electrons as incident particles of 500~keV. It is shown as  function of the scatter angle.}
\label{electron_dif_excit_fixE}
\end{figure}

 In Figure~\ref{electron_map_excit} is shown the relation between the incident energy and the scatter angle for $d\sigma/d\Omega$. As an example, in Figures~\ref{electron_dif_excit_fixangle} and \ref{electron_dif_excit_fixE} are shown the values of $d\sigma/d\Omega$ as function of the initial kinetic energy at 45$^o$ of scatter angle and the values of $d\sigma/d\Omega$ as function of the scatter angle for electrons of 500~keV.
 
 \subsubsection{Protons as incident particles}
\begin{figure}[htbp]
\begin{center}
\includegraphics[width=0.5\textwidth]{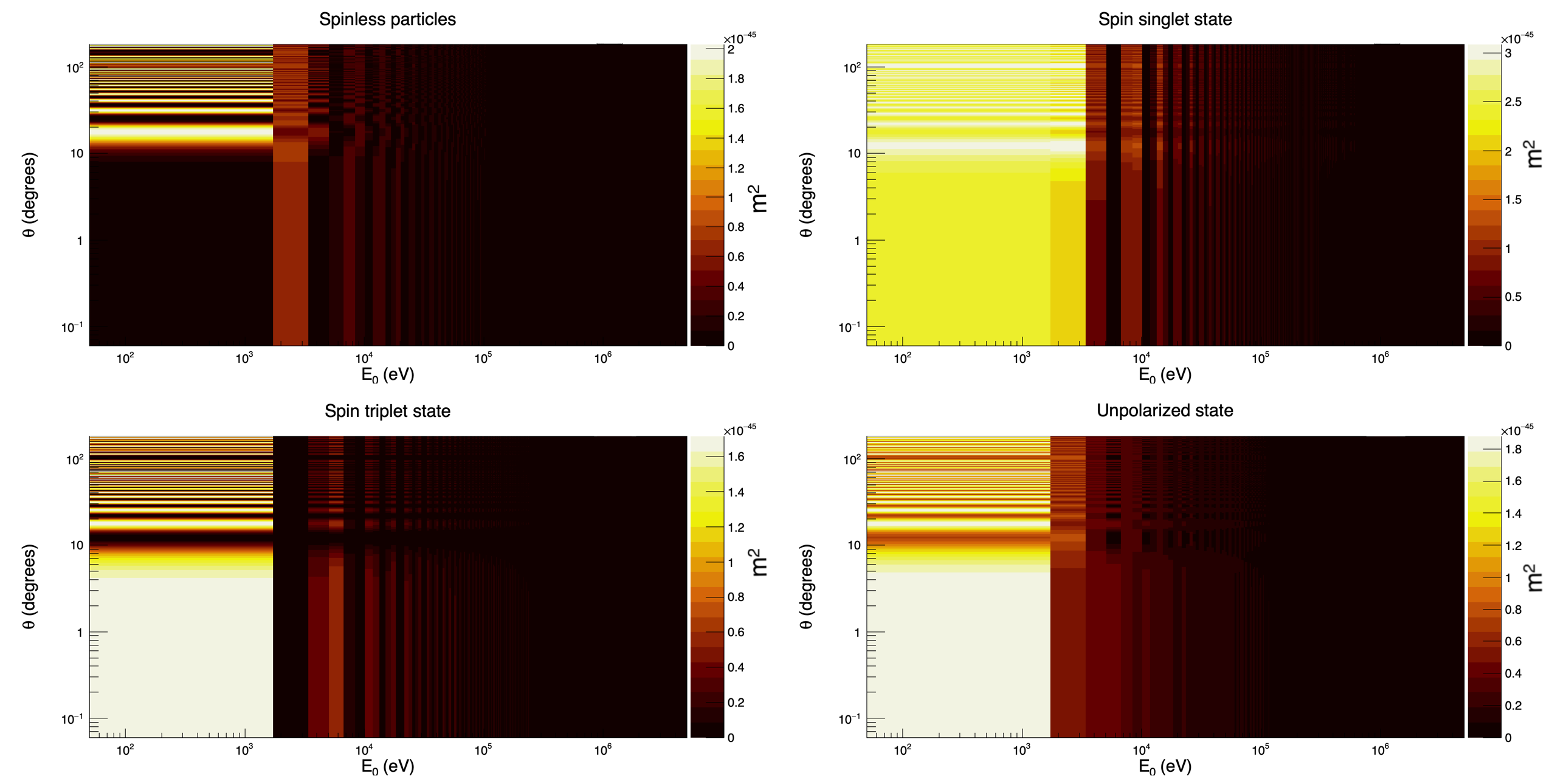}
\end{center}
\caption{$d\sigma/d\Omega$ values for the co-relation  between the scatter angle and kinetic energy for protons as incident particles, for the case of water molecule excitation. In the top left corner is shown the results for the case without spin, the top right corner is for spin singlet state, the bottom left corner shows the triplet case and the right bottom corner shows the results for the unpolarized state.}
\label{proton_map_excit}
\end{figure}

\begin{figure}[htbp]
\begin{center}
\includegraphics[width=0.5\textwidth]{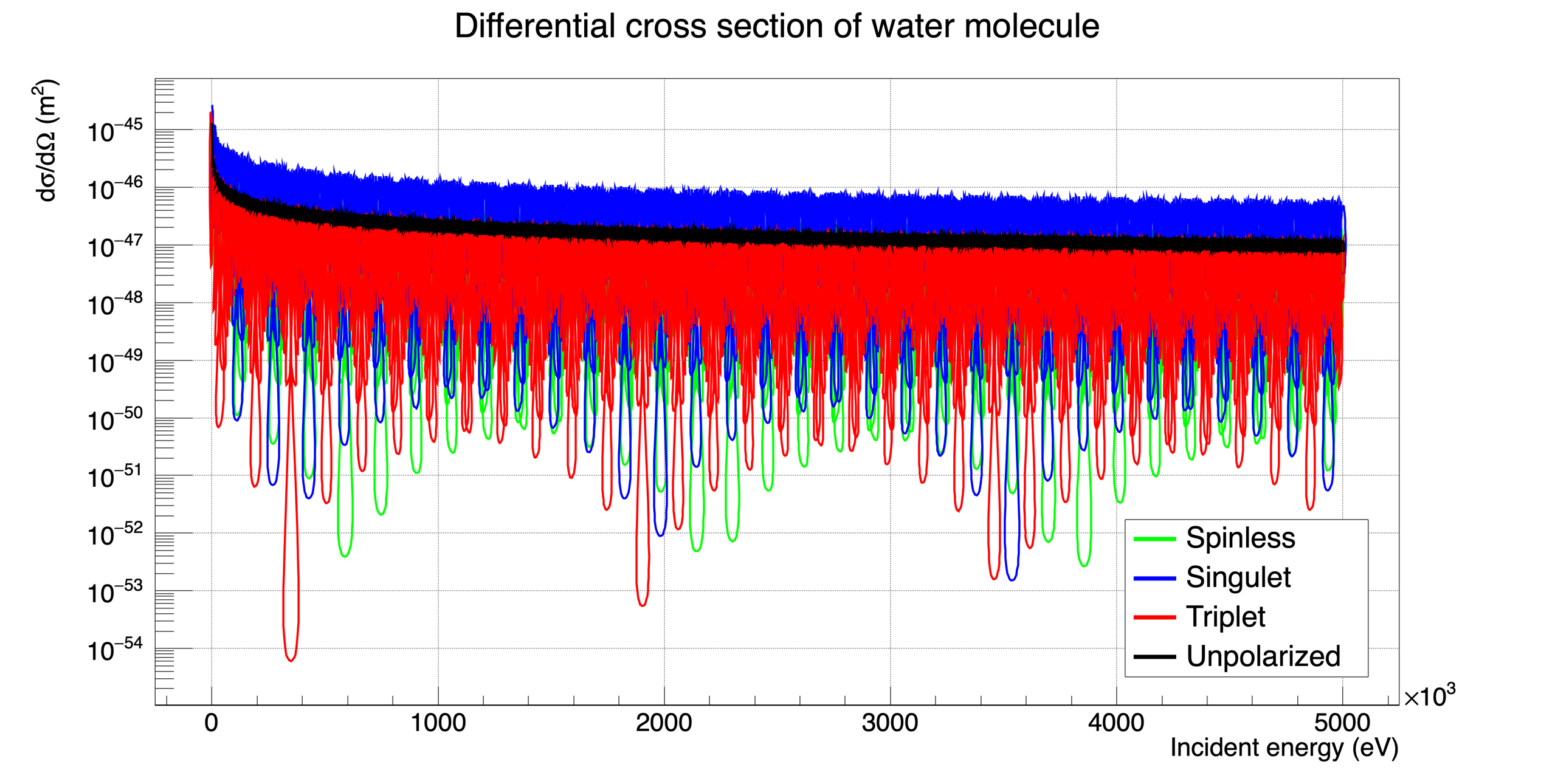}
\end{center}
\caption{Differential cross section for a scatter angle of 45$^o$, for the   excitation of water molecule for protons as incident particles. It is shown as a function of kinetic energy.}
\label{proton_dif_excit_fixangle}
\end{figure}

\begin{figure}[htbp]
\begin{center}
\includegraphics[width=0.5\textwidth]{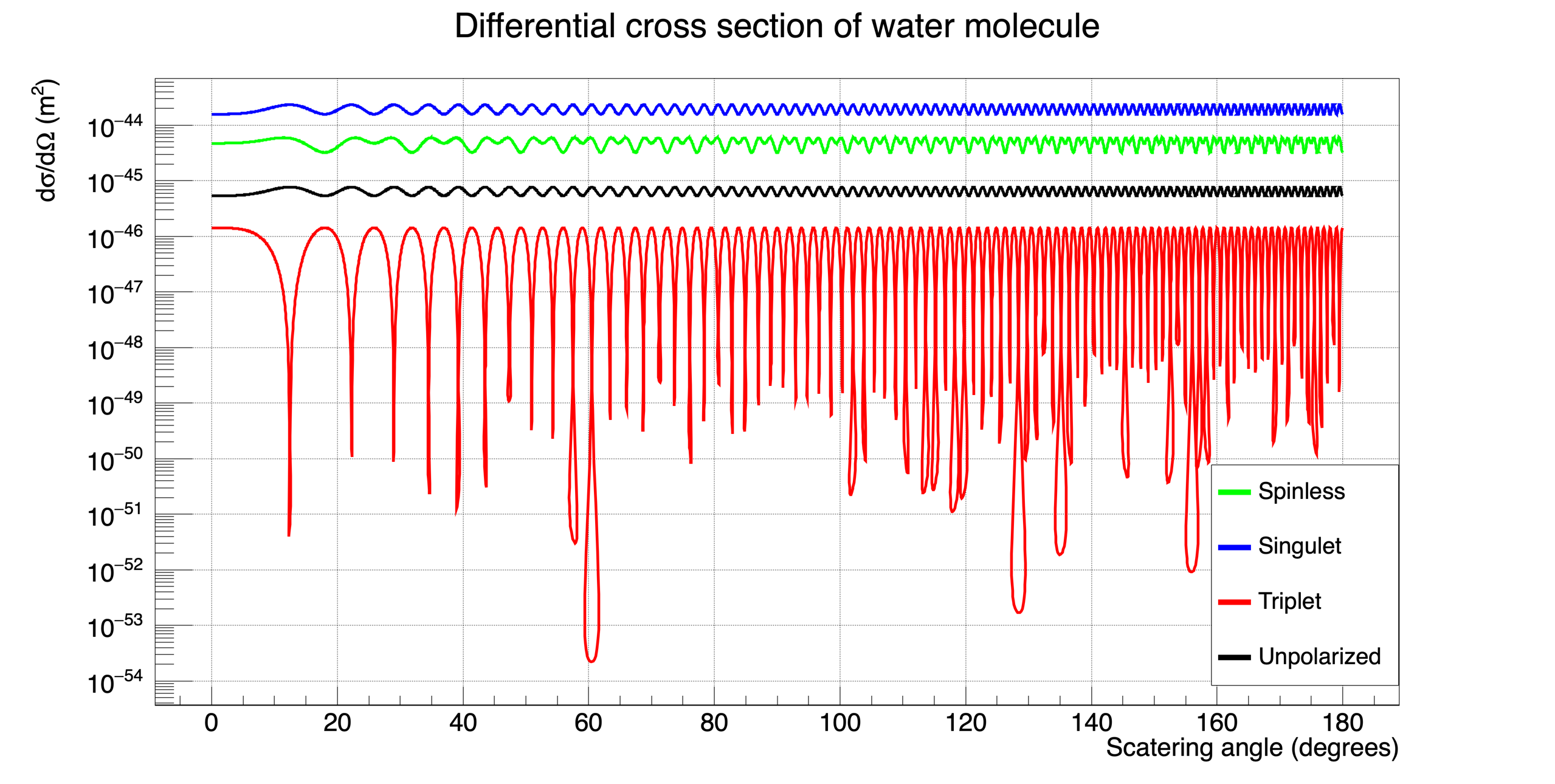}
\end{center}
\caption{Differential cross section for the   excitation of water molecule for protons as incident particles of 500~keV. It is shown as a function of the scatter angle.}
\label{proton_dif_excit_fixE}
\end{figure}

 When the incident particles are protons, it is obtain the phenomenon of scattering of identical particles, then, as an example, the co-relation between the scatter angle and the incident energy  is shown in Figure~\ref{proton_map_excit}. In Figures~\ref{proton_dif_excit_fixangle} and \ref{proton_dif_excit_fixE} are shown the values of $d\sigma_{fes}/d\Omega$ as function of the initial kinetic energy at 45$^o$ of scatter angle and the values of $d\sigma_{fes}/d\Omega$ as function of the scatter angle for electrons of 500~keV.
 
\subsection{Ionization of water molecule}
As it is described above, in this work the ionization of water molecule occurs when a proton is ejected, which leads to the production of the molecule $OH^-$. The ejected proton's wave function is as a free particle and it is not considered for the final state. Therefore, the final state is giving in terms of the scattered particle $\psi_s(\vec r)=e^{i\vec k_f \cdot \vec r}$ and the water molecule  wave function immediately after the ejected proton, which is $1/\sqrt{4\pi}$. The final state of the system is giving by $\displaystyle \Psi_f(\vec r)=\frac{1}{\sqrt{4\pi}}e^{i\vec k_f \cdot \vec r}$. Then, the differential cross section is giving by

\begin{equation}
    \begin{split}
        \displaystyle \frac{d\sigma_{ion}}{d\Omega}=&\frac{k_f}{k_i}|f(\theta,\phi)|^2\\
        =&\frac{k_f}{k_i}\Big|-\frac{\mu}{2\pi\hbar^2} \frac{1}{(4\pi)^{3/2}}\int \int \int e^{i\vec q_{ion} \cdot \vec r} \\ 
&\times \Big(\frac{2e^2}{r}-\frac{e^2}{|\vec r-\vec d_o^1|}-\frac{e^2}{|\vec r-\vec d_{o}^2|}\Big)  d\Omega_1 d\Omega_2 d^3r\Big|^2\\
    \end{split}
\end{equation}

where $\vec q_{ion}=\vec k_f-\vec k_i$. Also, for this case, the incident particle deposits energy, then  $k_f\neq k_i$. After solving the nine integrals, the differential cross section is giving by,

\begin{equation}\label{diff_ion}
  \displaystyle  \frac{d\sigma_{ion}}{d\Omega}=\frac{k_f}{k_i}\left(\frac{\mu e^{2}\sqrt{\pi}}{q_{ion}^{2}\hbar^2}\right)^2\left[1 - \frac{4}{\pi}\frac{sin(q_{ion}d_{o})}{q_{ion}d_{o}} \right]^2
\end{equation}

\subsubsection{Electrons as incident particles}
\begin{figure}[htbp]
\begin{center}
\includegraphics[width=0.5\textwidth]{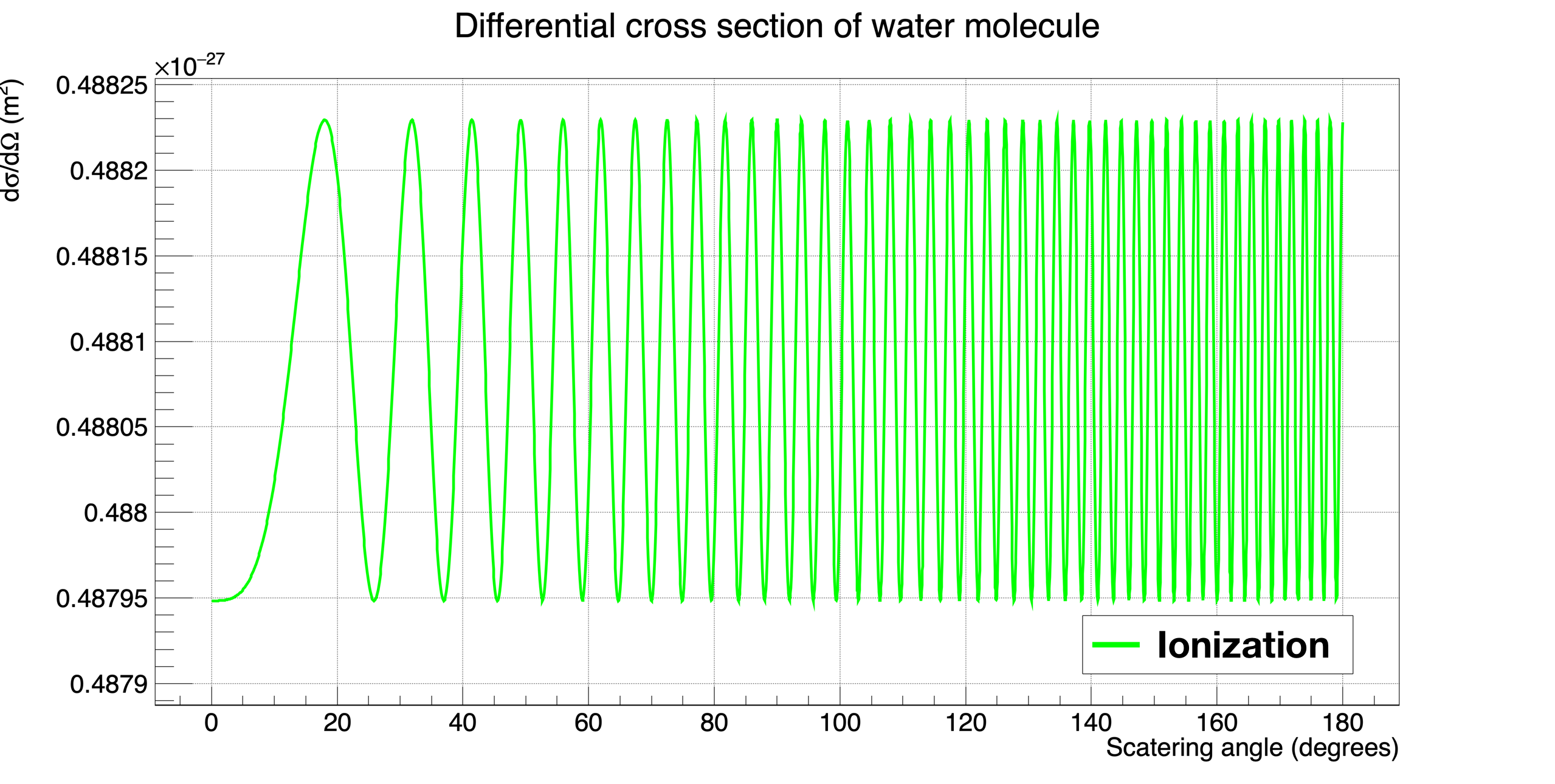}
\end{center}
\caption{Differential cross section as function of the scatter angle, for the excitation of water molecule for electrons as incident particles of 500~keV.}
\label{electron_dif_ion_fixangle}
\end{figure}

\begin{figure}[htbp]
\begin{center}
\includegraphics[width=0.5\textwidth]{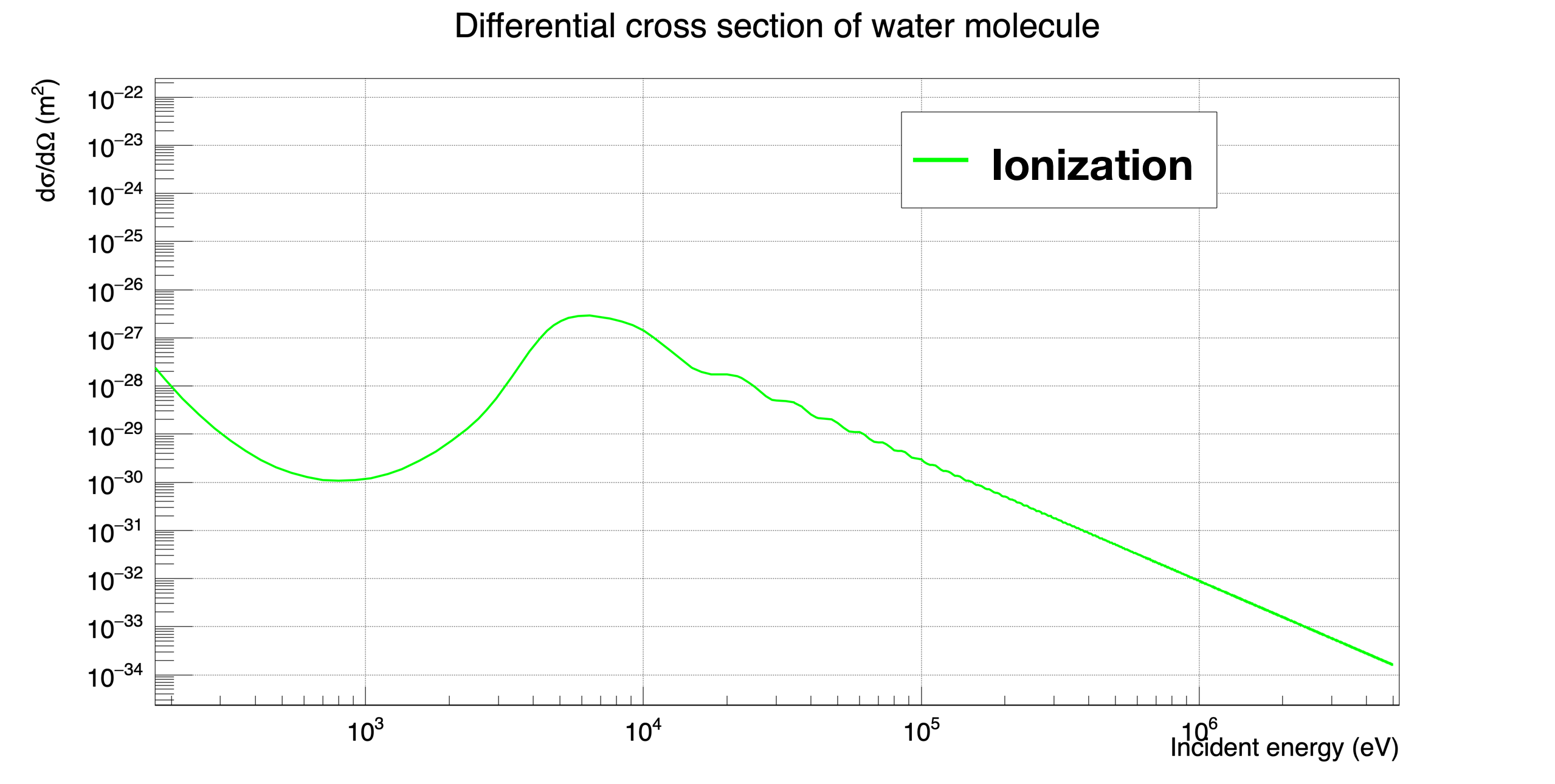}
\end{center}
\caption{Differential cross section for a scatter angle of 45, for the excitation of water molecule for electrons as incident particles as function of kinetich energy.}
\label{electron_dif_ion_fixE}
\end{figure}

\begin{figure}[htbp]
\begin{center}
\includegraphics[width=0.5\textwidth]{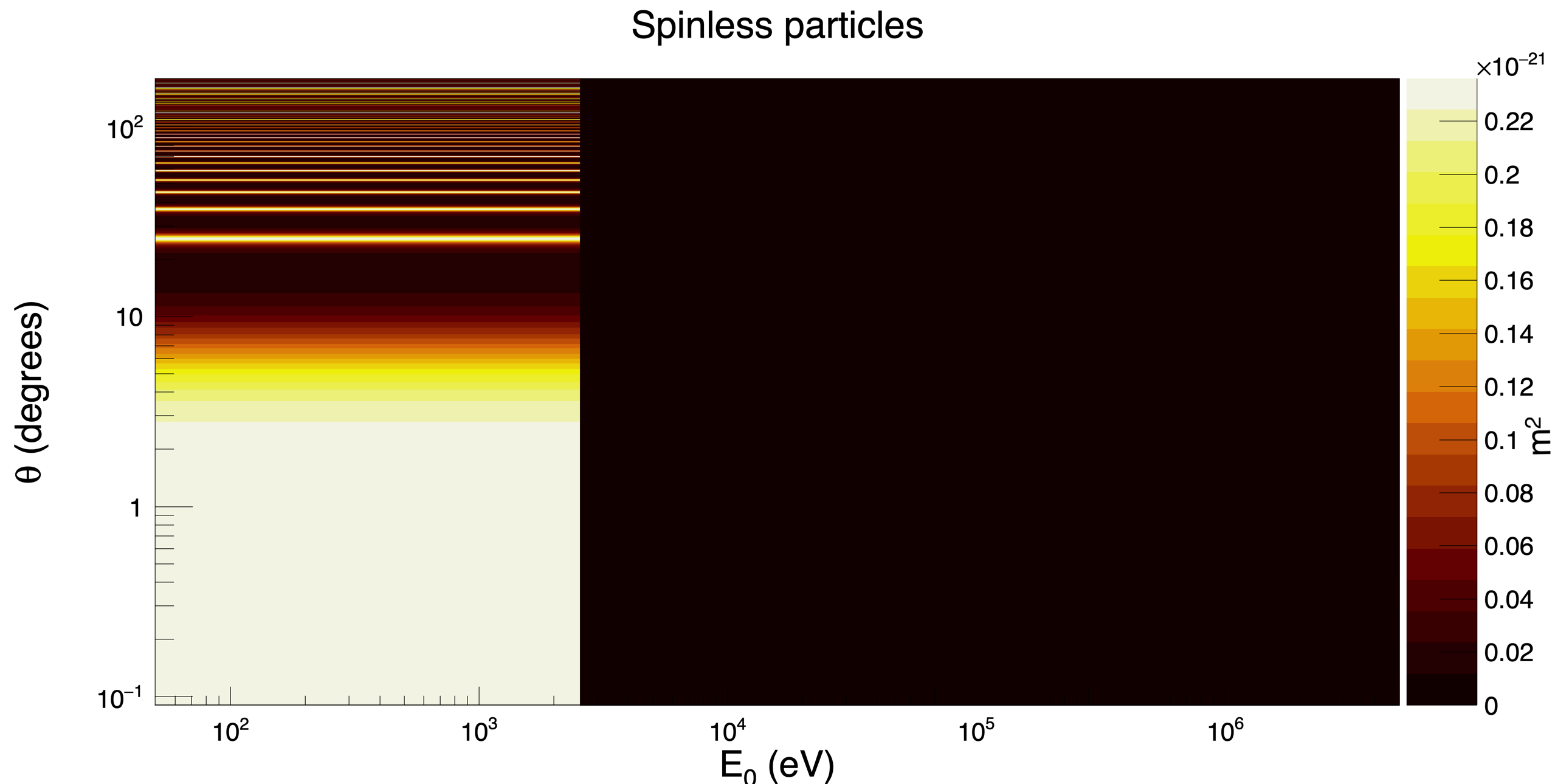}
\end{center}
\caption{Correlation for differential cross section between the scatter angle and kinetic energy for electrons, for the case of ionization of water molecule.}
\label{electron_dif_ion}
\end{figure}

In Figure~\ref{electron_dif_ion_fixangle} is shown the differential cross section as function of the angle, for electrons of 500~keV. In Figure~\ref{electron_dif_ion_fixE} is shown the differential cross section as function of the incident energy for a scatter angle of 45$^o$. Finally, in Figure~\ref{electron_dif_ion} is shown the correlation between the scatter angle and the incident energy.

\newpage
\subsubsection{Protons as incident particles}
\begin{figure}[htbp]
\begin{center}
\includegraphics[width=0.5\textwidth]{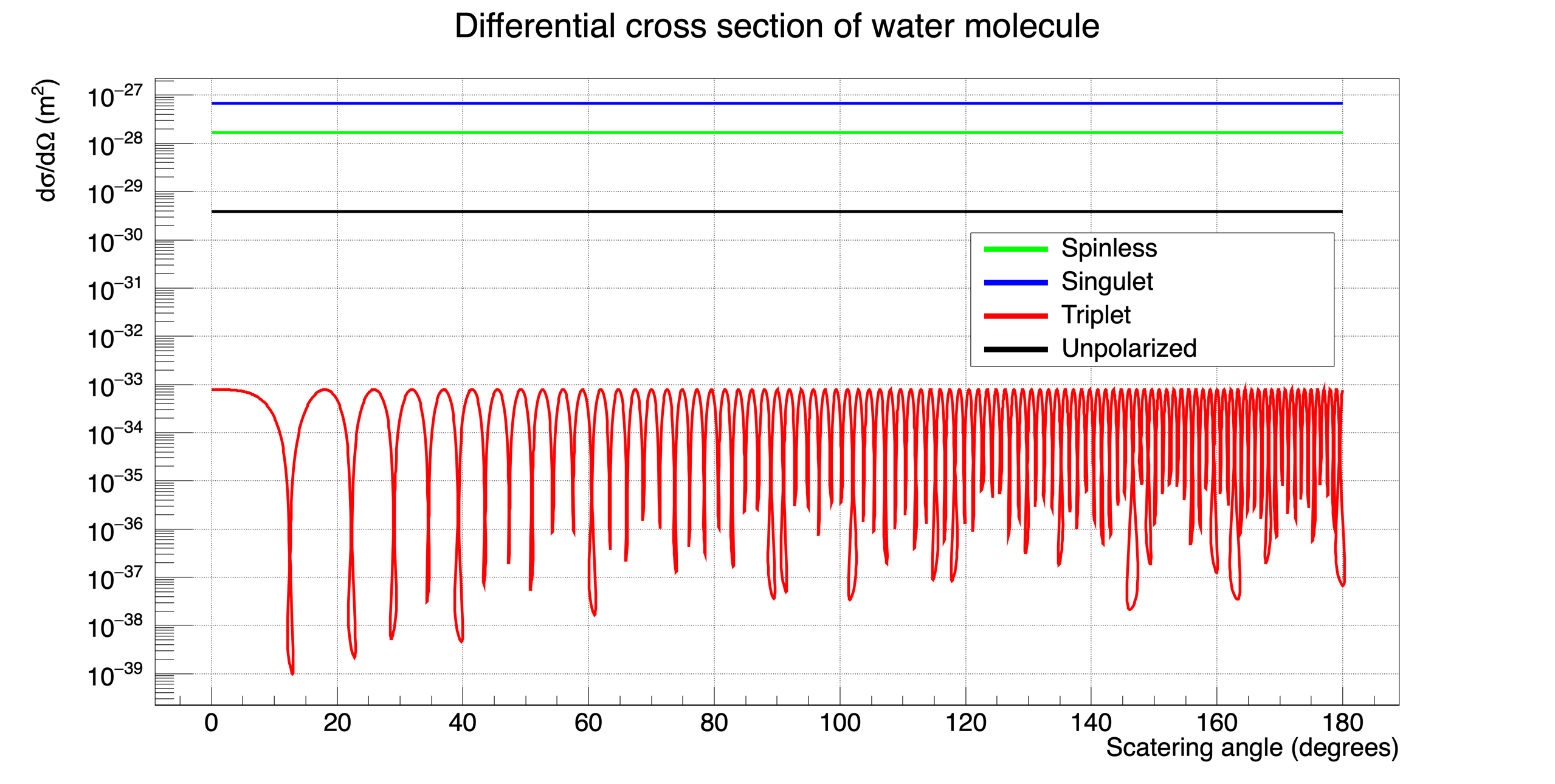}
\end{center}
\caption{Differential cross section for a scatter angle of 45$^o$ as function of kinetic energy, for the ionization of water molecule. When protons are incident particles.}
\label{proton_dif_ion_fixangle}
\end{figure}

\begin{figure}[htbp]
\begin{center}
\includegraphics[width=0.5\textwidth]{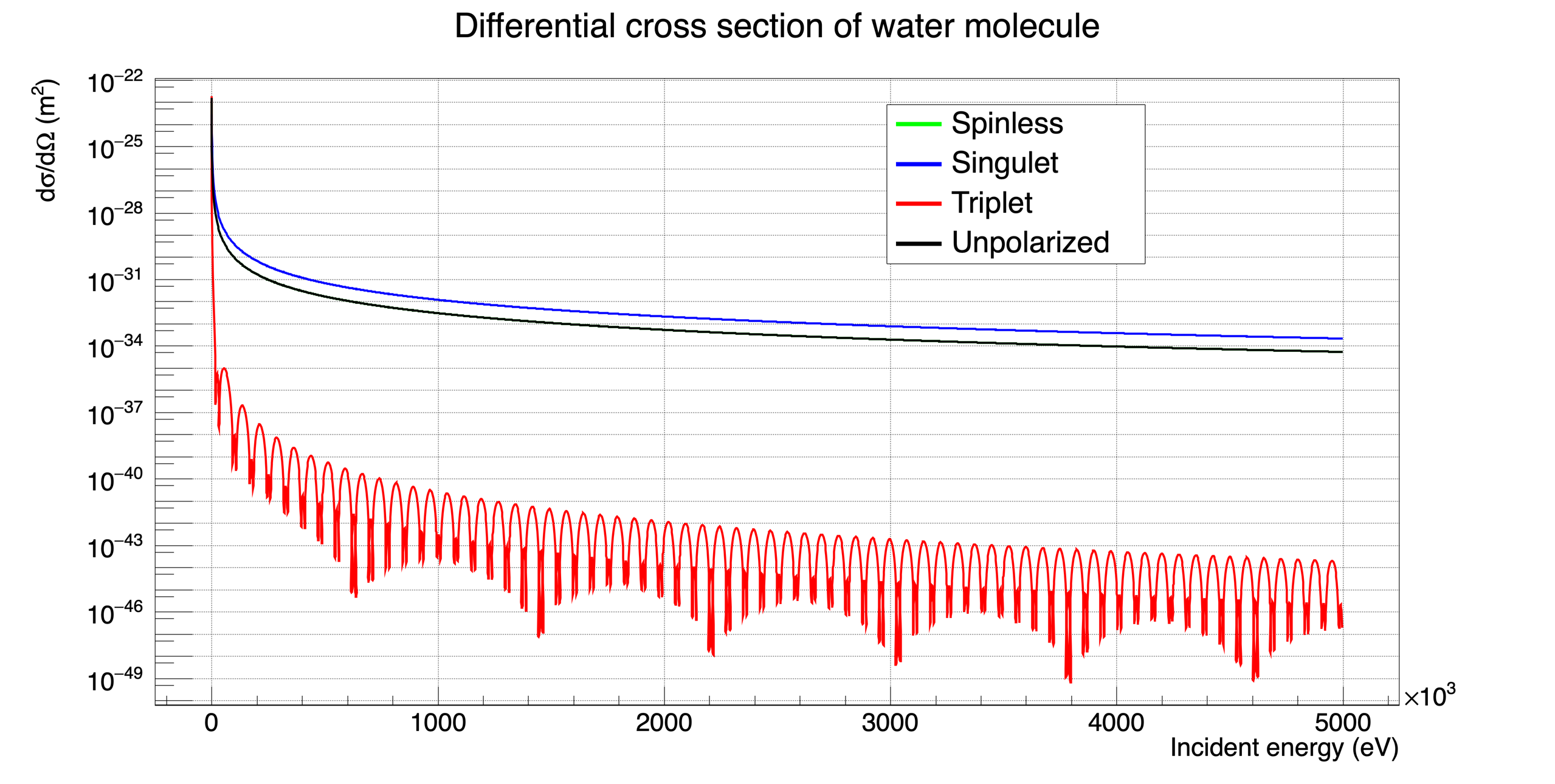}
\end{center}
\caption{Differential cross section for a scatter angle of 45$^o$ as function of kinetic energy, for the ionization of water molecule for protons as incident particles.}
\label{proton_dif_ion_fixE}
\end{figure}

\begin{figure}[htbp]
\begin{center}
\includegraphics[width=0.55\textwidth]{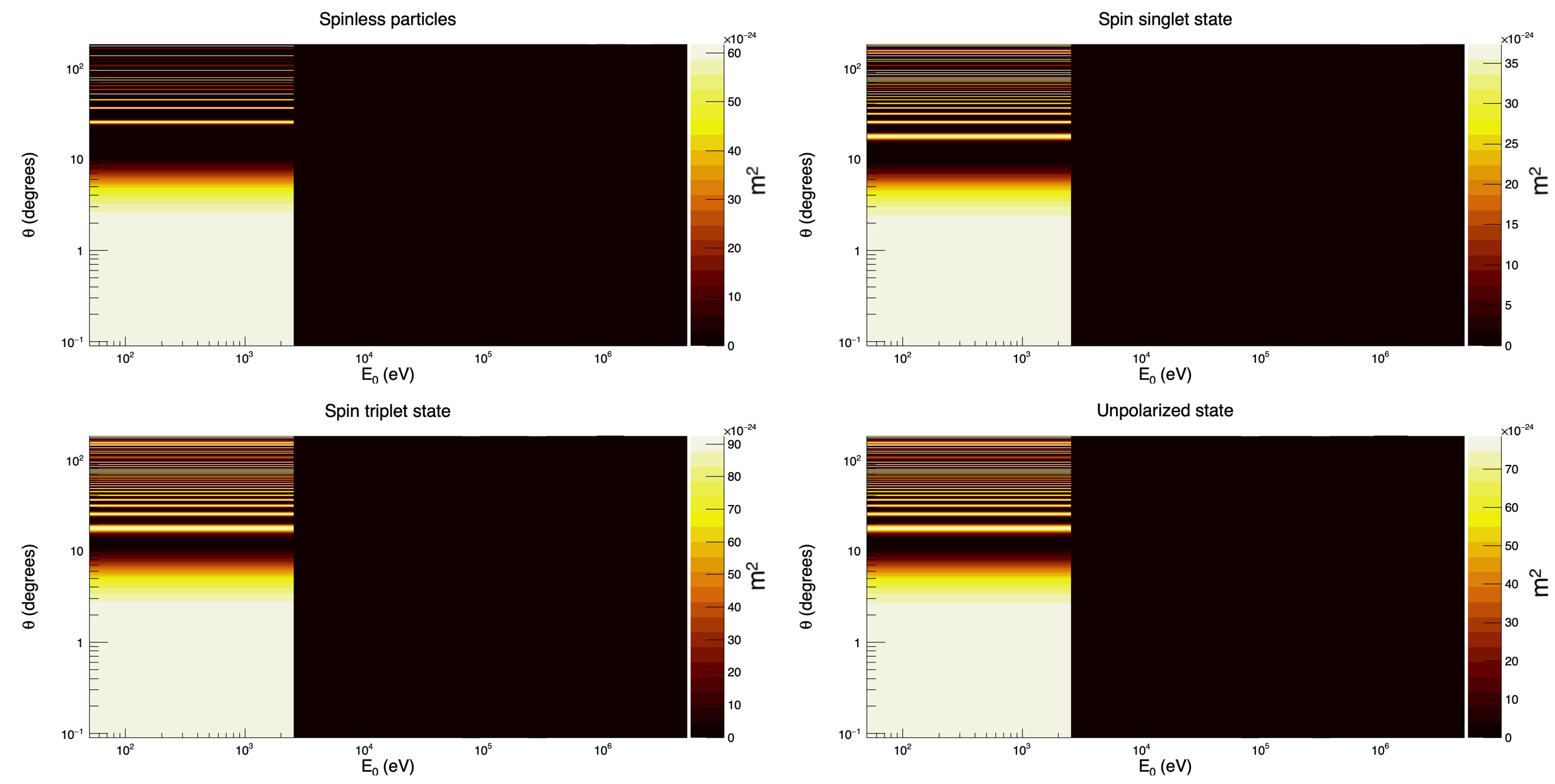}
\end{center}
\caption{Correlation for differential cross section between the scatter angle and kinetic energy for protons for the case of ionization of water molecule. In the top left corner is shown the results for the case without spin, the top right corner is for spin singlet state, the bottom left corner shows the triplet case and the right bottom corner shows the results for the unpolarized state.}
\label{proton_dif_ion}
\end{figure}

Once again, we have scattering of identical particles. In Figures~\ref{proton_dif_ion_fixangle}, \ref{proton_dif_ion_fixE} and \ref{proton_dif_ion} are shown the differential cross section as function of a fixed energy {500~keV}, fixed angle (45$^o$) and the correlation between the scatter angle and incident energy, respectively.

\section{Conclusions and discussions}\label{conclusions}
Make a model of the water molecule considering all its components (28 particles), it is practically impossible. Depending on the phenomenon to be studied, is where the model is selected. In some cases an electronic model is used, excepting the nuclei. In this work, it was used a model of three particles, where the two protons of the hydrogen model that make up the water molecule are the main particles. They move around the oxygen double negatively charged and unstructured, due to the electrostatic interaction, the distance between the three particles is constant. It was chosen this model, in order to calculate the differential cross section of the ionization of the water molecule, to obtain the $OH^-$ molecule, which consists of the water molecule losing a nucleus of the hydrogen atom.  To calculate the differential cross section, it was used the amplitude function described in quantum mechanics. In addition to the calculation of the ionization phenomenon, it was possible to calculate the  differential cross section for the cases of elastic collision and excitation of the water molecule. Where protons and electrons were considered  as incident particles. Due to the water model used, when the protons are considered as incident particles, there is a collision of identical particles, then, it was used the quantum theory of identical particles collision, to analyze this phenomena\\
The results obtained were that for the elastic collision $d\sigma/d\Omega$ decays as function of the incident energy, for both incident particles, as they are shown in Figs.~\ref{electron_dif_elastic_fixangle} and~\ref{proton_dif_elastic_fixangle}. However, when $d\sigma/d\Omega$ is considered as function of the scatter angle, the behavior is oscillatory, showing the phenomenon of interference, as it is shown in Figs.~\ref{electron_dif_elastic_fixE} and~\ref{proton_dif_elastic_fixE}. A correlation between the scatter angle, the incident energy and the $d\sigma/d\Omega$ are shown in Figs.~\ref{electron_map_elastic} and~\ref{proton_map_elastic}, where the black region does not represent that there is no data, but because of the order of magnitude it is very small with the other values.\\
For the case of excitation, $d\sigma/d\Omega$ oscillates as function of the particle incident, its  energy and the scatter angle, as they are shown in Figs.~\ref{electron_map_excit}, \ref{electron_dif_excit_fixangle},~\ref{electron_dif_excit_fixE},~\ref{proton_map_excit},~\ref{proton_dif_excit_fixangle} and~\ref{proton_dif_excit_fixE}. Then, it can begin to conclude that due to the phenomenon (elastic collision and excitation), the interference effect is more notorious. \\
Finally, for the ionization phenomena and according to the excitation phenomena, it can be concluded that depending on the type of particle and the dependence of $d\sigma/d\Omega$, the phenomenon of interference is presented.\\
In conclusion, these differential cross section results are in agreement with the scattering theory of quantum mechanics, in which the interference phenomenon is obtained. In a previous work~\cite{Houamer}, it is shown the interference phenomenon, for the ionization water molecule, unlike this work, which, the ionization was studied for the $OH^-$ production, by removing the proton from the nucleus of the hydrogen atom. Depending on the type of particle, its polarization and the phenomenon due to the collision, $d\sigma/d\Omega$ has a different shape and value. The  Eqs.~\ref{diff_elastic},~\ref{diff_exc} and~\ref{diff_ion} can be implemented in Monte Carlo simulations to help the microdosimetry studies.\\\\

\end{document}